\PassOptionsToPackage{super,sort&compress,comma}{natbib}
\documentclass[aps,twocolumn,superscriptaddress]{revtex4-2}

\usepackage{amsmath,braket,dsfont,amssymb,mathtools,bm,url,subcaption,setspace}
\usepackage{xcolor}

\definecolor{bordeaux}{RGB}{123, 27, 56}
\usepackage[colorlinks=true,linkcolor=bordeaux,bookmarks=false,citecolor=magenta,urlcolor=blue]{hyperref}
\usepackage[font=small,labelfont=bf,format=plain,justification=raggedright,singlelinecheck=false]{caption}
\usepackage{soul}
\usepackage{graphicx}
\usepackage{float}
\usepackage{tikz}
\usepackage{quantikz}
\usepackage{booktabs}
\usepackage{svg}
\usepackage[capitalise]{cleveref}
\crefname{figure}{FIG.}{FIGs.}
\crefname{section}{Sec.}{Secs.}
\usepackage{verbatim}

\newcommand{\F}{\texttt{Friend}}
\newcommand{\W}{\texttt{Wigner}}

\newcommand{\ketbra}[2]{\mathinner{|{#1}\rangle\langle{#2}|}}

\begin{document}

\title{Universal Quantum Computation in a Wigner's Friend Framework}

\author{Mart\'i Cladera-Rossell\'o}
\email{marti.cladera.rossello@univie.ac.at}
\affiliation{University of Vienna, Faculty of Physics, Vienna Center for Quantum Science and Technology (VCQ), Boltzmanngasse 5, 1090 Vienna, Austria}
\affiliation{University of Vienna, Vienna Doctoral School in Physics, Boltzmanngasse 5, 1090 Vienna, Austria}

\author{Sebastian Horvat}
\affiliation{University of Vienna, Department of Philosophy, Universitätsstraße 7, 1010 Vienna, Austria}

\author{Borivoje Daki\'c}
\affiliation{University of Vienna, Faculty of Physics, Vienna Center for Quantum Science and Technology (VCQ), Boltzmanngasse 5, 1090 Vienna, Austria}

\author{Lee A. Rozema}
\affiliation{University of Vienna, Faculty of Physics, Vienna Center for Quantum Science and Technology (VCQ), Boltzmanngasse 5, 1090 Vienna, Austria}

\author{Philip Walther}
\email{philip.walther@univie.ac.at}
\affiliation{University of Vienna, Faculty of Physics, Vienna Center for Quantum Science and Technology (VCQ), Boltzmanngasse 5, 1090 Vienna, Austria}
\affiliation{Christian Doppler Laboratory for Photonic Quantum Computer, Boltzmanngasse 5, 1090 Vienna, Austria}
\affiliation{Institute for Quantum Optics and Quantum Information (IQOQI) Vienna, Austrian Academy of Sciences, Boltzmanngasse 3, 1090 Vienna, Austria}

\date{\today}

\begin{abstract}
    {
    The Wigner's Friend thought experiment illustrates a core tension in quantum mechanics: an observer inside an isolated laboratory (the \F) measures a definite outcome, while an external observer (\W) describes the laboratory as a coherent superposition.
    Here we show that this discrepancy, usually studied as a conflict between descriptions, has a direct computational consequence. We namely propose a protocol by which \W\ can turn the \F’s laboratory into a universal \textit{quantum} computer, whereby the \F\ is restricted to apply simple \textit{classical} operations within the laboratory.
    We also experimentally implement a proof-of-principle demonstration of this protocol for single-qubit gates using single photons.
    By recasting the coherence between incompatible descriptions as a computational resource, we thus link the foundations of quantum measurement to quantum computation.
    }
\end{abstract}

\maketitle


 \section{Introduction}\label{sec:Intro}

The Wigner's Friend thought experiment, proposed in 1961 to highlight the measurement problem~\cite{wigner1961}, has recently come back into focus, through the use of modern quantum information tools, which allow for a more rigorous analysis~\cite{schmid2024,frauchiger2018,brukner,brukner2017,baumann2018,healey2018,zukowski2021,proietti2019,bong2020}.
In the original thought experiment, a \F\ performs a measurement of a quantum system inside an isolated laboratory, recording a definite outcome (from her point of view). 
However, from the perspective of an external agent, \W, who is powerful enough to treat the \F\ in her laboratory as a closed quantum system, the situation looks different.
\W\ describes the \F's state and the measured system as becoming entangled via the measurement interaction~\cite{zurek2003}. As such, the measurement outcome (and the \F\ herself) are left in a superposition, entangled with the quantum system.
Therefore, \W\ does not describe the laboratory as hosting a definite outcome, in seeming conflict with the \F's observations. 
This conflict, while implied by ordinary quantum theory, is surely puzzling, and is accordingly made sense of in different ways by different interpretations of the theory \cite{nurgalieva2021,rovelli1996,healey2018,baumann2018}.

A range of extensions to the above thought experiment have since been developed~\cite{schmid2024}, including nested observers~\cite{frauchiger2018}, no-go theorems on observer-independent facts~\cite{brukner,guerin2021},  on local friendliness~\cite{bong2020,wiseman2023}, and experimental tests of local observer independence~\cite{proietti2019,bong2020,laux2026}.
In all of these, the Wigner's Friend scenario is used to make the disagreement between the agents' descriptions explicit.
Here, instead, we propose a \textit{computational} consequence of the thought experiment. 
We show that \W\ can turn the \F’s laboratory---{which contains a rudimentary classical computer}---into a universal quantum computer even when the \F\ is restricted to applying simple bit permutations that cannot even achieve universal classical computation.
\W\ achieves this by sending the \F\ messages that are superpositions of instructions to apply different permutations.
The \F\ only ever reads a definite classical instruction and permutes bits; she never learns about the superposed message, which Wigner prepares outside the laboratory.
The resource allowing this computational power stems from \W's quantum-coherent manipulation of the entire lab, including the \F\ and her environment. Universal quantum computation would not be possible if information about the \F's operations were to leak outside of the lab.

\begin{figure}[t]
    \centering
    \includegraphics[width=\linewidth]{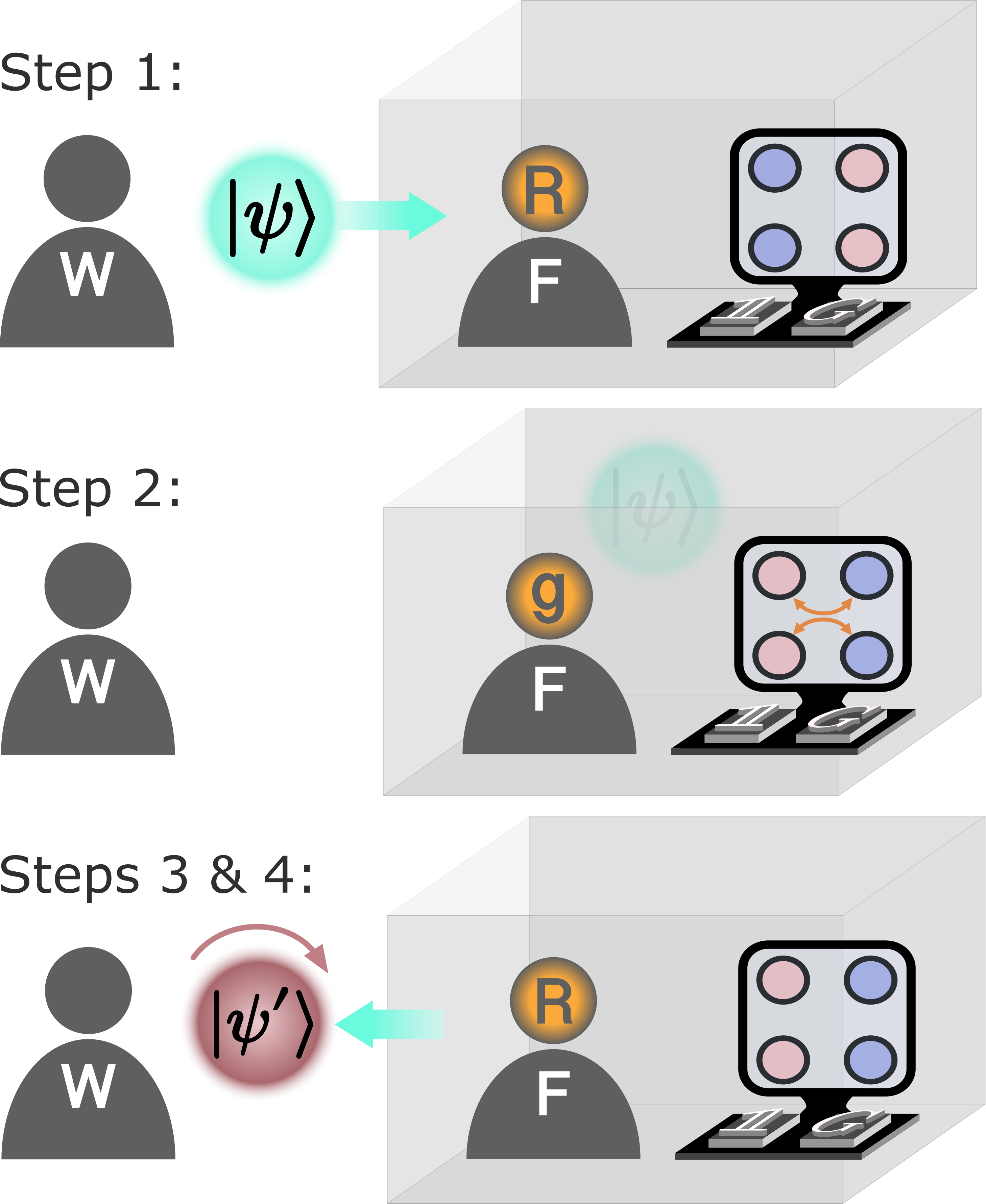}
    \caption{\textbf{{Simplified sketch of the protocol steps. }}
    \textit{Step 1:} \W\ (W) prepares the message $\ket{\psi}$ in a superposition state, and sends it to the laboratory, where the \F\ (F) waits, with her memory initially being in the ``ready state'' $R$. The only operations that the \F\ can do on her target register - illustrated as a $2\times2$ grid - are the identity operation or a permutation.
    \textit{Step 2:} The \F\ reads the message - with her memory recording its content - and applies the instructed operation on the target register.
    \textit{Step 3:} \W\ reverses the state of the whole lab, except for the target register: the \F's memory is thus again in state R. 
    In \textit{Step 4}, \W\ applies a unitary operation on the message and sends it back into the lab.\\
    Iterations of this procedure effectively generate unitary transformations on the target register.
    }
    \label{fig:simp_concept}
\end{figure}

Our result builds on recent work showing that quantum-coherently controlling a restricted set of classical operations, which by themselves are not universal for classical computation, is sufficient to achieve universal quantum computation~\cite{horvat2022}, building on a broader line of work on the coherent control of quantum operations~\cite{chiribella2013,araujo2014,procopio2015}. 
While that construction was formulated in an abstract control-target setting, here we embed it into a Wigner's Friend scenario where the \F\ acts as an observing mediator between \W's control message and a classical target register inside her laboratory.
More precisely, the \F\ is assumed to be located within a laboratory, where she has access to the aforementioned target system, on which computations are to be applied. \W\ is located outside of the laboratory, but still allowed to communicate with the \F, by exchanging messages through a quantum channel, taking the place of the control qubit. 
Our protocol then consists of the following steps:

\begin{enumerate}
    \item \W\ prepares a message to send into the laboratory, naming some of the classical operations that the \F\ is instructed to apply to her register of bits. This message may be prepared in a coherent superposition of different instructions (\cref{fig:simp_concept}{-\textit{Step 1}}).\\
    \item The \F\ receives the message, reads it, memorizes its content, and applies the corresponding operation on her target (\cref{fig:simp_concept}{-\textit{Step 2}}).
    \item The message is returned to \W, who then applies an ``erasure unitary operation'' to the whole laboratory except for the target system, thereby reversing the \F's state --including her memory -- to what it was before she had received the message (\cref{fig:simp_concept}{-\textit{Steps 3 \& 4}}).\\
    \item After this,  \W\ proceeds by either reading the message or applying a unitary operation on it (\cref{fig:simp_concept}{-\textit{Steps 3 \& 4}}), before again sending it back into the laboratory and repeating steps 1 to 3. 
\end{enumerate}

\noindent 
Remarkably, iterating this procedure allows \W\ to execute universal quantum computation on the target system, even though the \F\ only ever performs deterministic, non-universal classical operations. 
Because the \F's operations alone cannot even generate universal classical computation, {the protocol's computational power stems predominantly from Wigner's coherent control of the lab, including the \F\ and her memory.}

Notice that the technologically infeasible point of our protocol is step 3, which requires the possibility of quantum-coherently manipulating a macroscopic laboratory that contains a human being embedded in a complex environment.
Despite its (current) technological infeasibility, this possibility---{which has also been appealed to in previous work on Wigner's Friend} \cite{bong2020,evans2017}---is nevertheless a theoretically legitimate one, at least under the usual assumption that unitary quantum theory applies to arbitrarily complex physical systems~\cite{deutsch1985}.
Without this assumption our protocol could \textit{not} work, since records within the laboratory would essentially decohere the state of the target system. 

While the macroscopic implementation of our protocol can not currently be experimentally tested, here we implement a microscopic analogy thereof, using smaller quantum systems to demonstrate the use of our protocol to implement single-qubit gates.
In particular, we use a photonic system to implement the logical gate $R_X(\theta)=e^{-i\frac{\theta}{2}\hat{X}}$ at sixteen values of $\theta$ covering a full rotation about the logical $X$ axis, with average fidelity $\langle\mathcal{F}\rangle = 0.9885 \pm 0.0012$ and average purity $\langle\mathcal{P}\rangle = 0.9775 \pm 0.0023$ obtained by maximum-likelihood quantum state tomography~\cite{hradil1997quantum}. 
In our experiment, the target system and the \F\ are encoded in different degrees of freedom of a single photon~\cite{englert2001}, and are thus not physically separate systems. 
Nevertheless, our experiment does demonstrate a realization of the formal protocol structure sketched above. Whether this structure can be scaled to progressively larger systems, however, remains an open experimental challenge. 

\section{Results}
\subsection{Framework}\label{sec:theor_frame}

Aside from \W, three systems are needed for our protocol: a target system made out of $2n$ qubits (bits from the \F's perspective), a quantum message, and the \F. 
The latter three will enter the quantum-mechanical descriptions. 
\W\ will, on the contrary, not be modeled explicitly, as he is an external agent that conducts experiments on the laboratory and on the message. 


\subsubsection{State-spaces}
\label{sec:spaces}
\noindent\emph{Target system and classical operations.} The \F\ is assumed to be located inside a laboratory, where she has access to a target system made out of $2n$ qubits.
She is, however, only allowed to execute a limited range of operations on the target, amounting to mere permutations in a fixed basis. 
This makes said qubits effectively behave as classical bits.
More precisely, let the so-called ``classical basis'' be spanned by vectors $\{\ket{\vec{x}}\equiv\ket{x_1\dots x_{2n}}\,,\ x_i\in\{0,1\}\}$. We assume that the \F\ is only capable of implementing operations drawn from the following set
\begin{equation}
    \mathcal{G} = \{\mathds{1}\}\cup\bigcup_{l=1}^{n}\left\{G_1^{(l)},G_2^{(l)}\right\},
    \label{eq:gset}
\end{equation}
where $G_i^{(l)}\equiv\mathds{1}^{\otimes(2l-2)}\otimes G_i\otimes\mathds{1}^{\otimes(2n-2l)}$ acts non-trivially only on the $l$-th qubit pair as
\begin{equation}
\begin{split}
&G_1\ket{x_lx_{l+1}}=\ket{(x_l\oplus1)x_{l+1}},\\ &G_2\ket{x_lx_{l+1}}=\ket{x_l(x_{l+1}\oplus x_l)}.
    \label{eq:gates}
\end{split}
\end{equation}
Each element of $\mathcal{G}$ is a permutation of the classical basis: either a NOT on the first qubit pertaining to a pair, or a CNOT on the pair, or the identity.
Relative to the \F's operations, the $2n$ qubits thus behave as if they were $2n$ classical bits. 
Note that NOTs and CNOTs alone cannot generate even the AND operation. The \F\ can accordingly not even implement universal \textit{classical} computation, let alone quantum computation.
Therefore, all of the computational power of our protocol will have to come from outside the laboratory.\\



\noindent\emph{Message.} The message is a quantum system that \W\ uses to instruct the \F\ which operation she is to execute within the laboratory. 
The \F\ will read the message by measuring it in the computational basis and from her point of view observe a definite outcome.
The state space $\mathcal{H}_M$ associated to the message will accordingly correspond to the space of possible operations that the \F\ is capable of carrying out in the laboratory:
\begin{equation}
    \mathcal{H}_M=\mathrm{span}\left\{\ket{g}_M : g\in \mathcal{G}\right\},
    \label{eq:hm}
\end{equation}
Each state $\ket{G_i^{(l)}}_M$ carries the instruction to execute operation $G_i^{(l)}$, whereas $\ket{\mathds{1}}_M$ carries the instruction to do nothing at all. 
Note that no restriction is placed on the quantum operations that \W\ is allowed to implement on the message outside of the laboratory, though only a few limited single-qubit gates will already suffice. \\


\noindent\emph{Friend.} The \F\ remains inside the lab, and reads messages sent by \W, telling her which operation to perform on the target system. 
As usual in Wigner-Friend scenarios, we will model the \F\ as a memory that keeps a record of the operations that she has implemented. 
We thus choose to model the state-space $\mathcal{H}_F$ associated to the \F\ in analogy with the state-space associated to the message:
\begin{equation}
    \mathcal{H}_F=\mathrm{span}\left(\{\ket{\mathcal{R}}_F\}\cup\{\ket{g}_F : g\in \mathcal{G}\}\right).
    \label{eq:hf}
\end{equation}
Each state $\ket{G_i^{(l)}}_F$ represents the \F\ after having read and executed instruction $G_i^{(l)}$; $\ket{\mathds{1}}_F$ represents the \F\ after having read the instruction to do nothing, and having done nothing; finally, the \textit{ready state} $\ket{\mathcal{R}}_F$ represents the \F\ before having received any message at all. 



\subsubsection{Operations}

\noindent\emph{Read and execute.} In our protocol, \W\ sends quantum messages into the laboratory (step 1), where the \F\ reads it and follows the instructions (step 2). 
We model this latter step as a sequence of two transformations. 
First, the reading is described by a \textit{coupling unitary} $C$, which transitions the \F's memory from its ready-state into a record of the read instruction:
\begin{equation}
    C\,\ket{g}_M\ket{\mathcal{R}}_F=\ket{g}_M\ket{g}_F,\qquad g\in \mathcal{G}.
    \label{eq:coupling}
\end{equation}
Next, the \F\ executes the instruction by applying the corresponding gate to the target system: if she reads that operation $g$ is to be executed, she applies this $g$ on the target system. 
The net effect of this two-step procedure is 
\begin{equation}
    \ket{g}_M\ket{\mathcal{R}}_F\ket{\phi}\;\longmapsto\;\ket{g}_M\ket{g}_F\,g\ket{\phi},\qquad g\in \mathcal{G},
    \label{eq:action}
\end{equation}
where $\ket{\phi}$ is an arbitrary state of the target system. Note that \cref{eq:coupling} specifies $C$ only on a specific list of input states---those in which the \F\ is in a ready-state---thereby underspecifying the total unitary.\\



\noindent\emph{The erasure unitary.} After step 2, the message and the \F's memory are correlated. 
If this correlation were to survive, the \F's memory would effectively amount to a which-path record that destroys the interference between different instruction branches. 
In order to maintain coherence, \W\ must therefore reset the \F's memory to its ready-state (step 3). 
More precisely, after the \F\ reads the message, executes its instruction, and sends the message outside of the laboratory, \W\  implements an ``erasure-unitary'' $E$ that acts as follows:
\begin{equation}
    E\,\ket{g}_M\ket{g}_F=\ket{g}_M\ket{\mathcal{R}}_F,\qquad g\in \mathcal{G}.
    \label{eq:E_def}
\end{equation}
This requires coherent quantum control over the entire laboratory, similar to the additional elements of the extended Wigner's Friend scenarios~\cite{deutsch1985,proietti2019,bong2020}; we will return to this in \cref{sec:discussion}.



\subsubsection{Logical Qubit encoding}\label{ssec:log_enc}
As introduced in Sec. \ref{sec:spaces}, the \F's target system consists of $2n$ target qubits.
In our protocol, these will effectively encode $n$ ``logical qubits'' with basis vectors given by: 
\begin{equation}
    \ket{0}_L\equiv\tfrac{1}{\sqrt2}(\ket{10}-\ket{11}),\qquad \ket{1}_L\equiv\tfrac{1}{\sqrt2}(\ket{00}-\ket{01}),
    \label{eq:log_enc}
\end{equation}
where the classical basis vectors $\ket{xy}$ correspond to pairs of adjacent target qubits. 
The $l$-th logical qubit will accordingly be encoded in the above two-dimensional subspace of target qubits $(2l-1,2l)$, for any $l=1,...,n$.
The logical qubit states cannot be prepared by the \F\ alone, since they are given by superpositions of classical basis states. Nevertheless, as we show in \cref{met:prep}, they can be prepared with \W's help, via a procedure similar to the general protocol presented below.

Note that the \F's gates---which amount to mere permutations in the classical basis---effectively act as Pauli gates in the logical basis:
\begin{equation}
    G_1^{(l)}\ket{y}_L=\ket{y\oplus1}_L,\qquad G_2^{(l)}\ket{y}_L=-(-1)^y\ket{y}_L,
    \label{eq:gate_log}
\end{equation}
Thus, in the logical basis: $G_1^{(l)}\equiv\hat{X}$ and $G_2^{(l)}\equiv-\hat{Z}$ .

\subsection{The protocol}\label{sec:protocol}
\begin{figure}[t]
    \centering
    \includegraphics[width=\linewidth]{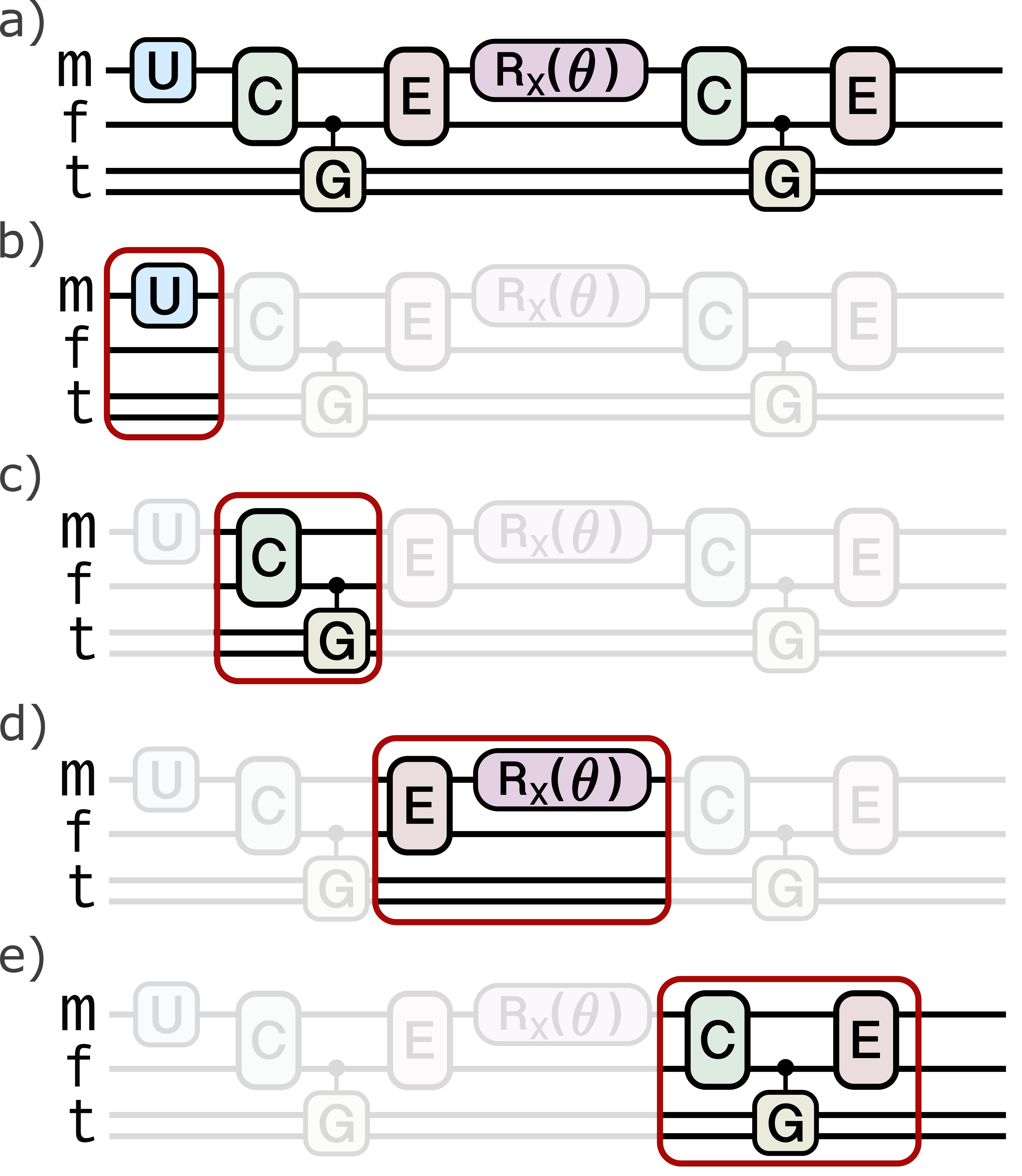}
    \caption{\textbf{{Quantum circuit of the protocol.  }}
    \textbf{a)} Complete quantum circuit on the message (m), memory (f) and target (t), whose two wires are the pair of bits the \F\ always acts on in the classical basis. 
    Initially, the state of the message is an arbitrary $\ket{m}_M$ state, the memory is in $\ket{\mathcal{R}}_F$, and the target qubit is already in a given logical state $\ket{\varphi}_L$.
    \textbf{b)} \textit{Step 1}: \W\ prepares the message $\ket{+}_M=(\ket{\mathds{1}}_M+\ket{G}_M)/\sqrt{2}$, by acting on in it with a unitary, $U$.
    \textbf{c)} \textit{Step 2}: the coupling $C$ takes her memory from $\ket{\mathcal{R}}_F$ to a record of what she read, $\ket{g}_F$.
    In one branch it becomes $\ket{\mathds{1}}_F$ and she leaves the register alone; in the other it becomes $\ket{G}_F$ and applies the classical operation.
    \textbf{d)} \textit{Steps 3 and 4}: the erasure $E$ resets the \F's memory to $\ket{\mathcal{R}}_F$.
    \W\ then applies $R_X(\theta)$ to the message, encoding the desired rotation angle.
    \textbf{e)} \textit{Steps 2 and 3} are repeated, leaving the message in the original state after \W's initialisation, $\ket{+}_M$, the memory in state $\ket{\mathcal{R}}_F$, and having applied $e^{-i\theta G/2}$ to the input target state,  which for $G=G_1$ is the logical rotation $R_X(\theta)$.
    }
    \label{fig:concept}
\end{figure}
\begin{figure*}[t]
    \centering
    \includegraphics[width=\textwidth]{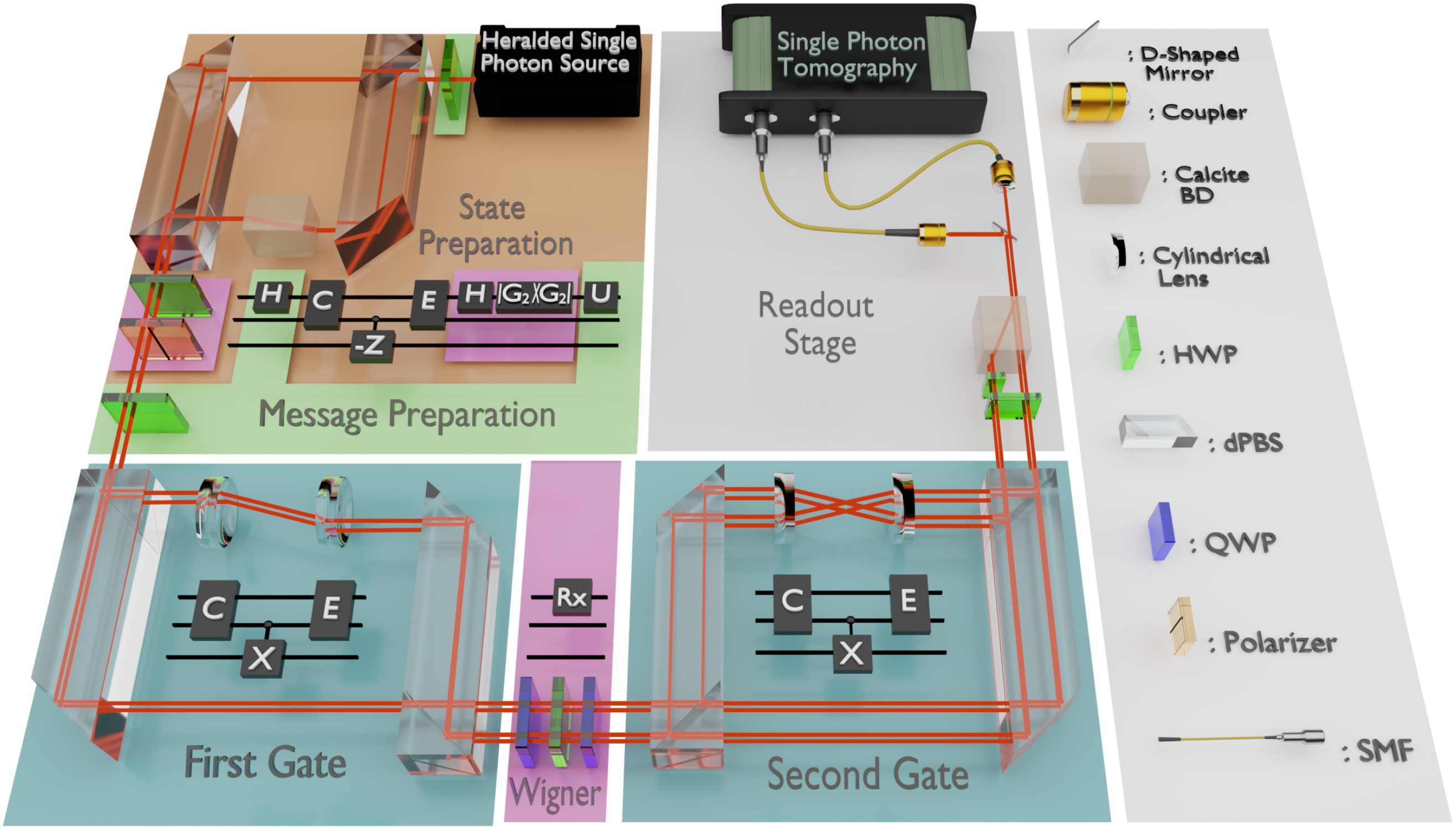}
    \caption{\textbf{{Illustration of the experimental setup.}} The experiment uses heralded single photons at $1550$~nm. Polarisation encodes the message $M$, a $2\times2$ grid of spatial modes encodes the classical basis, and the path relative to the double polarising beam splitters (dPBS) encodes the \F's memory. Wigner prepares the message by using a half-waveplate to modify the polartisation (green). State preparation (orange) converts the input into the logical state $\ket{0}_L$ using a calcite beam displacer, waveplates and a polariser. In pink we have the action corresponding to step 4. Each gate stage (blue) implement steps 2 and 3: the entry dPBS is the coupling $C$, a pair of cylindrical lens telescope applies the permutation $G_1=\hat{X}$ to the reflected arm, carrying the record $\ket{G_1}_F$, and the exit dPBS is the erasure $E$. Between the two rounds \W\ (pink) applies the rotation $R_X(\theta)$ to the message using a quarter-, half-, quarter-waveplate gadget. The output is analysed on the readout stage (grey) by performing single-photon tomography. The legend at right identifies the optical components.}
    \label{fig:setup}
\end{figure*}

We will now show how combining the elements introduced above enables the implementation of single-qubit rotations $e^{-i\frac{\theta}{2}X}$ and $e^{i\frac{\theta}{2}Z}$, as well as two-qubit entangling gates $e^{-i\frac{\theta}{2}X\otimes X}$ and $e^{-i\frac{\theta}{2}Z\otimes Z}$, for arbitrary angles $\theta$. 
Since these gates form a universal set, this will imply that our protocol allows \W\ to execute any quantum gate on the $n$-logical-qubit subspace of the target system 
\footnote{It is to be noted that our protocol structure resembles the exponential-swap construction of Ref.~\cite{Lau}.}.
Implementing each basis gate requires one full iteration of the four-step scheme announced in \cref{sec:Intro}, followed by a second repetition of Steps 2 and 3, as will be detailed in what follows - see also \cref{fig:concept} for the circuit that represents the protocol.

\subsubsection{Single-qubit gates}\label{ssec:single}
Let $\ket{\psi}_L$ be an arbitrary state of the $l$-th logical qubit. To implement a rotation generated by $G_i^{(l)}$ ($i=1,2$), \W\ initializes the message in the superposition $\ket{+_i^{(l)}}_M \equiv \tfrac{1}{\sqrt{2}}(\ket{\mathds{1}}_M + \ket{G_i^{(l)}}_M)$ and sends it into the laboratory (Step 1). The \F\ then reads the message and applies the corresponding classical permutation on her target register (Step 2), after which \W\ applies the erasure unitary $E$, restting the \F's memory to $\ket{\mathcal{R}}_F$ (Step 3). Up to this point, the joint state of the message, the \F, and the logical target evolves as
\begin{align}
\xrightarrow{\text{Step 1}}&\ket{+_i^{(l)}}_M \ket{\mathcal{R}}_F \ket{\psi}_L \nonumber \\
\xrightarrow{\text{Step 2}}& \frac{1}{\sqrt{2}}\left(\ket{\mathds{1}}_M \ket{\mathds{1}}_F \ket{\psi}_L + \ket{G_i^{(l)}}_M \ket{G_i^{(l)}}_F\, G_i^{(l)}\ket{\psi}_L\right) \nonumber \\
\xrightarrow{\text{Step 3}}& \frac{1}{\sqrt{2}}\left(\ket{\mathds{1}}_M \ket{\psi}_L + \ket{G_i^{(l)}}_M\, G_i^{(l)}\ket{\psi}_L\right)\ket{\mathcal{R}}_F . \label{eq:round1}
\end{align}
With the \F's memory factored out, the message and the target register remain entangled. In Step 4, \W\ applies the single-qubit rotation $R_i^{(l)}(\theta) = e^{-i\frac{\theta}{2}X_M}$ to the returned message, corresponding to a rotation by angle $\theta$ around the $X$ axis of the two-dimensional message subspace spanned by $\{\ket{\mathds{1}}_M, \ket{G_i^{(l)}}_M\}$.
He then sends it back into the laboratory for a second read-execute-erase round (repeating Steps 2 and 3). As it can be easily verified, this second round decouples the message and leaves the joint system in the final state
\begin{equation}
\ket{+_i^{(l)}}_M\,\ket{\mathcal{R}}_F\;e^{-i\frac{\theta}{2}G_i^{(l)}}\ket{\psi}_L .
\label{eq:final}
\end{equation}
Since $G_1^{(l)} \equiv X$ and $G_2^{(l)} \equiv -Z$ on the $l$-th logical subspace (\cref{eq:gate_log}), choosing $i=1$ implements $e^{-i\frac{\theta}{2}X}$ while choosing $i=2$ implements $e^{i\frac{\theta}{2}Z}$ on the logical qubit, leaving both the message and the \F's memory restored to their initial states.

\subsubsection{Two logical qubits and universality}\label{ssec:two}
It is simple to notice that, by replacing gate $G^{(l)}_i$ by any other gate $G$ that satisfies $G^2=\mathds{1}$, our protocol can implement the according rotation $e^{i\frac{\theta}{2}G}$.
Now, since $(G^{(l)}_iG^{(k)}_i)^2=\mathds{1}$, an analogous protocol can also be used to generate two-qubit gates $e^{{-}i \frac{\theta}{2} G^{(l)}_iG^{(k)}_i}$ on the logical qubit-pair $(l,k)$. 
To implement these the state-spaces of the message and the \F\ simply need to be expanded to include states that correspond to pairs of operations: e.g. $\ket{G_i^{(l)},G_i^{(k)}}_M$ is the state of the message that contains the instruction to implement $G_i^{(l)}G_i^{(k)}$, whereas $\ket{G_i^{(l)},G_i^{(k)}}_F$ is the state of the \F\ after reading this message and executing its instruction. 

Running the steps from before, with the message instead initially prepared in $\ket{+_i^{(l,k)}}_M \equiv \big(\ket{\mathds{1}^{(l)},\mathds{1}^{(k)}}_M+\ket{G_i^{(l)},G_i^{(k)}}_M\big)/\sqrt{2}$, thus yields output state
\begin{equation}
    \ket{+_i^{(l,k)}}_M\,\ket{\mathcal{R}}_F\;e^{-i\frac{\theta}{2}G_i^{(l)}\otimes G_i^{(k)}}\ket{\psi}_L,
    \label{eq:two_qubit}
\end{equation} 
thereby implementing $e^{-i \frac{\theta}{2} X \otimes X}$ or $e^{-i\frac{\theta}{2}Z \otimes Z}$ on the logical qubit-pair $(l, k)$. These entangling gates, together with the prior single-qubit gates, make universal quantum computing possible. 


\subsection{Experimental demonstration}\label{sec:results}

\begin{figure}[t]
    \centering
    \includegraphics[width=0.95\linewidth]{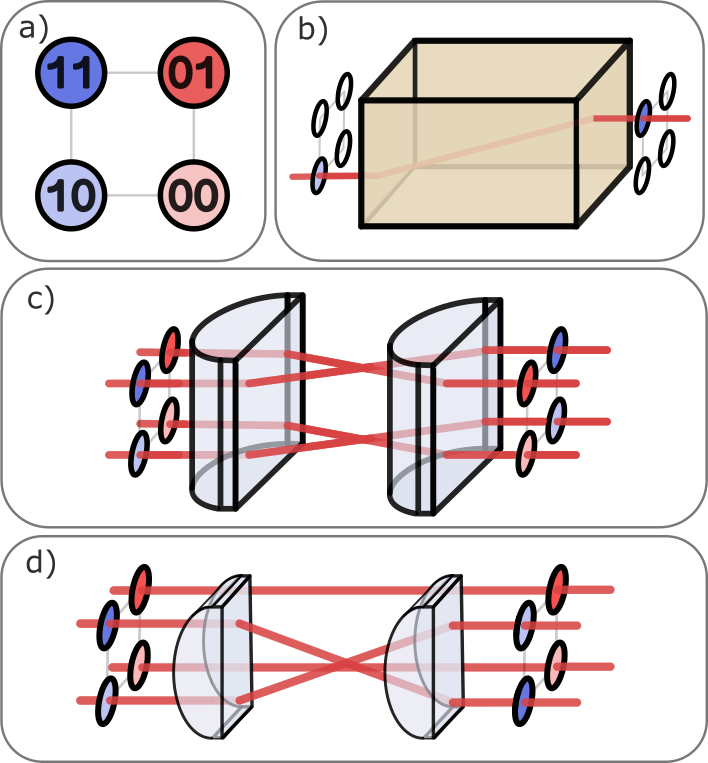}
    \caption{\textbf{Experimental encoding and gates.}
    Optical implementation of the classical gates on the spatially encoded target. \texttt{a)} Bit encoding: the classical basis $\{\ket{00},\ket{01},\ket{10},\ket{11}\}$ is mapped onto the four spatial modes of a $2\times2$ grid. \texttt{b)} The $G_2$ ($-\hat{Z}$) gate used during state preparation, implemented with a calcite beam displacer that shifts the input $\ket{10}$ mode vertically onto $\ket{11}$. \texttt{c)} The $G_1$ ($\hat{X}$) gate used in the logical gate stage, implemented with a $1{:}1$ cylindrical-lens imaging system whose optical axis is aligned vertically with the centre of the grid, exchanging the two columns and mapping $\ket{x_1 x_2}\mapsto\ket{(x_1\oplus 1)x_2}$. \texttt{d)} The same $1{:}1$ imaging system with its axis rotated to horizontal, and placed so that only the first column passes through the lenses and the bits $\ket{10}\leftrightarrow\ket{11}$ (the rows in the first-column) are swapped, giving $G_2$ ($-\hat{Z}$) on the full grid.}
      \label{fig:gates}
\end{figure}

\begin{figure*}[t]
    \centering
    \includegraphics[width=\linewidth]{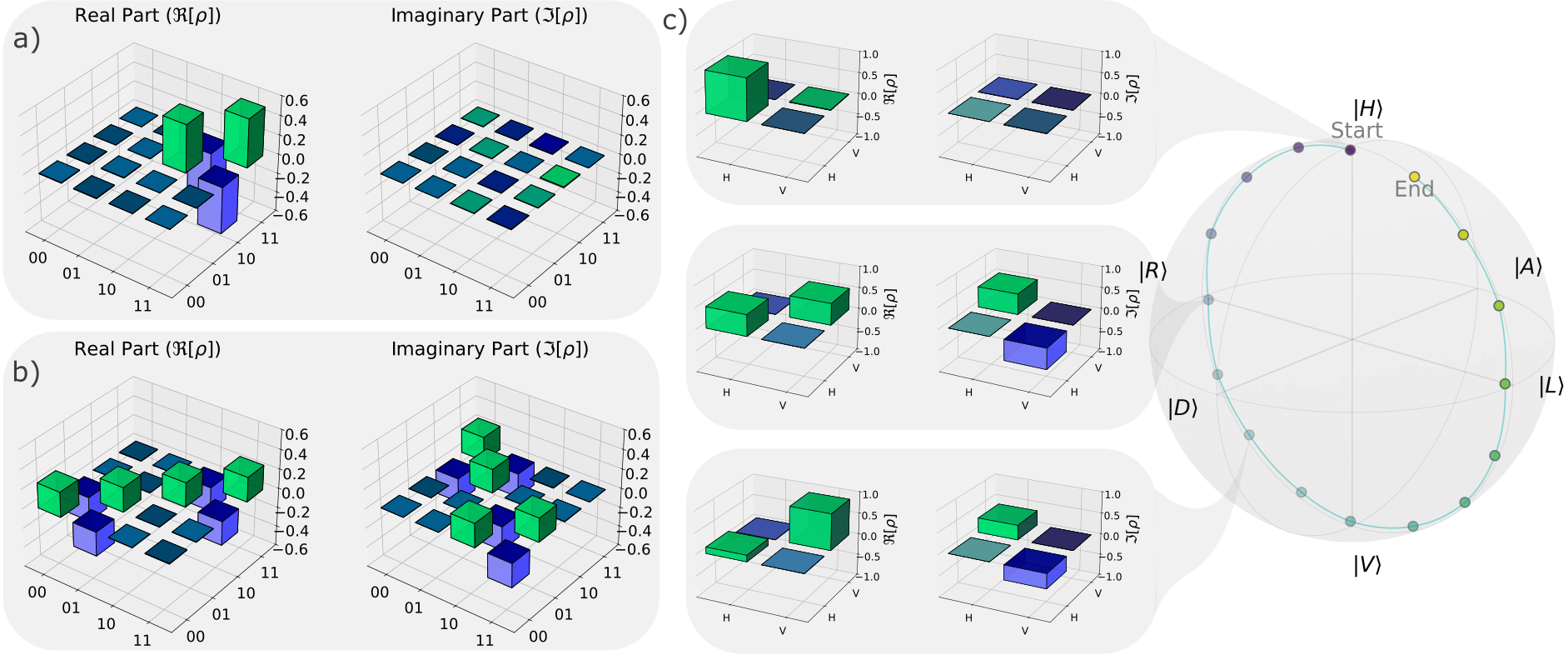}
    \caption{%
       \textbf{Reconstructed experimental density matrices and Bloch sphere evolution.} \textbf{a)} Reconstructed density matrix in the classical basis of the input state $\ket{0}_L$, after state preparation and prior to the protocol implementation, with $\mathcal{F}=0.9932(12)$. \textbf{b)} . Reconstructed target qubit density matrix in the classical basis for the target state at $\theta/2=\pi/4$, the reconstruction yields a fidelity of $\mathcal{F}={0.986(2)}$. \textbf{c)} Bloch sphere evolution of target logical qubit under different angles. For each angle we conducted a tomography, where we mapped the logical basis to polarisation basis. Three of the tomographies are shown for angles: $\frac{\theta}{2}=\left\{0,\frac{\pi}{4},\frac{3\pi}{8}\right\}$, with respective fidelities of $\mathcal{F}=\left\{0.9894(19),0.986(2),0.988(7)\right\}$.}
      \label{fig:data}
\end{figure*}

We will now experimentally implement the single-qubit gate $e^{-i\frac{\theta}{2}\hat X}$ using our protocol.
That is, we set $n{=}1$ and $G=G_1^{(1)}$.
Our experiment has two stages, shown in \cref{fig:setup}: \W, following the steps in \cref{met:prep}, first prepares $\ket{0}_L$ from a classical bit string (orange), and then runs steps 1-4 followed by another round of step 2 and step 3. Step 1 is shown in the green region, steps 2 and 3 are represented in the blue areas, while step 4 is shown in the pink area.
All of the required state spaces are encoded in a single heralded photon at $1550$~nm.\\

\noindent\emph{Encoding.}
Each system is encoded in a different degree of freedom (DOF) of the same photon.
The ``classical'' two-bit target system is encoded in a $2\times2$ grid of spatial modes.
A single photon in each mode thus represents a two-bit string as labeled in \cref{fig:gates}{\color{bordeaux}a}.
The \F's allowed classical operations are permutations of these spatial modes.
Experimentally, gates $G_1$ and $G_2$ are implemented with cylindrical-lens telescopes to permute modes as illustrated in \cref{fig:gates}{\color{bordeaux}c} and {\color{bordeaux}d}, respectively.
For the state preparation protocol (\cref{met:prep}), the \F\ implements $G_2$.
However, since for this step the photon enters the experiment in a well-defined spatial mode (i.e. in the ``classical state'' $\ket{10}$), we use a different implementation of $G_2$ shown in \cref{fig:gates}{\color{bordeaux}b}.
In particular, we use a calcite beam displacer to shift the mode upwards by $2.7$~mm. This allows us to precisely set the row spacing, which will be essential for our final verification of the protocol, wherein the spatial modes must be coherently recombined (readout stage of \cref{fig:setup}).

\W's message is encoded in the polarisation of the same photon with $\ket{H}\leftrightarrow\ket{\mathds{1}}_M$ and $\ket{V}\leftrightarrow\ket{G}_M$.
To read the message, the \F\ must couple it to her memory.
This is accomplished with a double polarising beam splitters (dPBS) as follows.
The \F's memory is another path degree of freedom of the photon.
Before the dPBS, the memory is in the ready state $\ket{\mathcal{R}}_F$, as there is only a single path and no record of any instruction.
Upon entering the dPBS $\ket{H}$ is transmitted and $\ket{V}$ reflected.
Thus the entire grid is transmitted or reflected.
This is the coupling $C$ of \cref{eq:coupling}.
The transmitted arm encodes $\ket{\mathds{1}}_F$. Thus, this arm is empty and does nothing.
On the other hand, the reflected arm represents $\ket{G}_F$, in which the permutation $G$ is placed.
The absence or presence of $G$ thus implements the controlled-$\ket{G}$ gates of Fig. \ref{fig:concept}.  
After this, a second dPBS recombines the two arms into one path without measuring which one the photon took, which implements the erasure $E$ of \cref{eq:E_def}.
Unlike a standard PBS, a dPBS sends its transmitted and reflected outputs out parallel to each other, allowing the interferometer to close with fewer mirrors than a conventional one.
This matters because mirrors can introduce incoherence between the two branches that the protocol must keep coherent.
In practice the setup remains stable without active stabilisation for at least 30 minutes, which is longer than a full characterization measurement (Supplementary Information, SI~\ref{met:adddata}).

In summary, \W\ only acts on the polarization of the photon (the message). The \F\ reads the polarization by coupling it to spatial modes using dPBSs.  She cannot otherwise act on the polarization, she is only able to implement permutations on the spatial modes.\\

\noindent\emph{State preparation.}
The photon starts in $\ket{H}\ket{10}$, with the message reading $\ket{\mathds{1}}_M$.
\W\ applies a Hadamard to the message with a HWP at $22.5^\circ$ and sends the photon through an interferometer whose reflected arm holds the beam displacer, where the \F\ executes $G_2$ as $\ket{10}\mapsto\ket{11}$ (\cref{fig:gates}{\color{bordeaux}b}).
A second Hadamard and a polariser transmitting $\ket{V}$ then leave the target in $\ket{0}_L$, with success probability $1/2$.
The coherence between $\ket{10}$ and $\ket{11}$ comes directly from the coherence of \W's message: the \F\ simply coupled the message to her additional spatial modes (memory) by displacing the photon based on its polarization.\\

\noindent\emph{Gate.}
After the state preparation, \W\ acts on the message state again, using a HWP at $67.5^\circ$ to rotate $\ket{V}$ to $\ket{D}$ and prepares $\ket{+}_M\ket{0}_L$.
In the first gate stage, the reflected arm holds two cylindrical lenses of $80$~mm focal length in a $1{:}1$ inverting telescope, axes vertical and centred on the grid.
The telescope swaps the two columns, $\ket{x_1x_2}\mapsto\ket{(x_1\oplus1)x_2}$, which is $G_1$ (\cref{fig:gates}{\color{bordeaux}c}).
On the logical qubit it takes $\ket{0}_L$ to $\ket{1}_L$, so after the first exit dPBS
\begin{equation}
    \ket{D}\ket{0}_L\;\longrightarrow\;\tfrac{1}{\sqrt2}\big(\ket{H}\ket{0}_L+\ket{V}\ket{1}_L\big).
\end{equation}
Between the two gate stages, \W\ rotates the polarisation with a quarter-, half-, quarter-wave plate sequence, $R_X(\theta)=\mathrm{QWP}(90^\circ)\,\mathrm{HWP}(\theta/4)\,\mathrm{QWP}(90^\circ)$.
The angle $\theta$ is set by the angle of a single HWP.
With $c=\cos\frac{\theta}{2}$ and $s=\sin\frac{\theta}{2}$, the state becomes
\begin{equation}
    \tfrac{1}{\sqrt2}\Big[\ket{H}\big(c\ket{0}_L-is\ket{1}_L\big)+\ket{V}\big(c\ket{1}_L-is\ket{0}_L\big)\Big].
    \label{eq:rot_exp}
\end{equation}
The second gate stage is a copy of the first.
It flips the logical qubit in the $\ket{V}$ branch only, which makes the two branches equal and leaves
\begin{equation}
    \ket{D}\otimes\big(c\ket{0}_L-is\ket{1}_L\big)=\ket{+}_M\,e^{-i\frac{\theta}{2}\hat X}\ket{0}_L,
    \label{eq:exp_out}
\end{equation}
as predicted by \cref{eq:final}.
No measurement appears anywhere in this stage, so it is deterministic.
The \F's optics never see $\theta$.
Both gate stages are identical at every angle, and the only element that changes is \W's HWP.

\noindent\emph{Readout.}
To verify the action of the gate, we step outside of the framework, and perform tomography on the logical qubit space as follows.
To read out the logical qubit we map it onto polarisation, $\ket{0}_L\to\ket{H}$ and $\ket{1}_L\to\ket{V}$.
A HWP across the whole grid takes $\ket{D}$ to $\ket{H}$, a second HWP rotates the right column alone to $\ket{V}$, and a calcite beam displacer merges the two columns.
Each row then carries the logical state in its polarisation, and the two rows carry the same state up to a global phase (\cref{met:tomo}).
We analyse the top row with a QWP, a HWP and a PBS.
Both PBS outputs are coupled into single-mode fibres and detected with superconducting nanowire single-photon detectors (PhotonSpot), in coincidence with the herald; these elements are not shown in \cref{fig:setup}.
The photon pairs come from type-0 spontaneous parametric down-conversion in a $30$~mm PPKTP crystal inside a Sagnac loop, pumped in one direction only by a $775$~nm continuous-wave laser so that the pair carries no polarisation entanglement.
For each tomography setting we acquire data $10$~s, resulting in approximately $1550$ coincidences per measurement.
The density matrices are reconstructed by maximum-likelihood estimation~\cite{hradil1997quantum} with Monte Carlo simulations used to estimate uncertainties.
Note that the readout writes the logical qubit onto the message's polarisation, and it assumes the message has returned to $\ket{D}$, as in \cref{eq:exp_out}.
Without resetting the message, errors would appear in the reconstructed logical state, so the fidelities below jointly test the gate and the resetting of the message.\\

\noindent\emph{Results.}
Our tomography results for different gates are shown in \cref{fig:data}.
In \cref{fig:data}{\color{bordeaux}a} we display the reconstructed density matrix of the full two-qubit target system in the classical basis gives $\mathcal{F}=0.9932\pm0.0012$ when $\theta=0$.
In this case, the expected output state is $\ket{0}_L$, which is apparent from the off-diagonal terms between $\ket{10}$ and $\ket{11}$.
In \cref{fig:data}{\color{bordeaux}b} we set $\theta/2=\pi/4$, and again perform full two-qubit tomography, observing coherence between the logical states $\ket{0}_L$ and $\ket{1}_L$, as expected.
The fidelity with $\ket{\theta/2}=\cos(\theta/2)\ket{0}_L-i\sin(\theta/2)\ket{1}_L$ is $\mathcal{F}=0.986\pm0.002$.
To visualize this more easily, in panel {\color{bordeaux}c} we show the logical states directly.
Scanning $\theta/2\in[0,\pi)$ in steps of $\pi/16$, and performing a tomography at each step.
The Bloch vectors follow the great circle that $R_X(\theta)$ traces through the $Y$--$Z$ plane (\cref{fig:data}{\color{bordeaux}c}).
Averaged over all sixteen angles, the state fidelity is $\langle\mathcal{F}\rangle=0.9885\pm0.0012$ and the purity is $\langle\mathcal{P}\rangle=0.9775\pm0.0023$.
A check that the bottom row reproduces the top-row state is given in \cref{met:tomo} and SI \ref{met:adddata}.
Two losses are built into the scheme.
The preparation passes half of the photons, and the readout uses one row out of two, so the overall success probability is at most $1/4$.

The residual infidelity is consistent with technical imperfections of the setup, such as polarisation mixing from the finite extinction ratio of the dPBS and spatial-mode aberrations introduced by the cylindrical-lens telescope.
Imperfections of this kind can leave the two arms of the interferometer partially distinguishable after the exit dPBS, so that some information about which operation was performed survives the erasure.
In SI~\ref{met:resource} we introduce a simple model of this situation, in which such information leaks into an environment. 
The environment is then left in one of two states, where $V$ is the real part of their overlap and $V=1$ for a perfect erasure.
Within this model, the applied operation becomes an incoherent mixture of the target gate ($e^{-i\frac{\theta}{2}\hat X}\ket{0}_L$) and a rotation by the same angle with opposite sign, $e^{+i\frac{\theta}{2}\hat X}\ket{0}_L$.
Imperfections in the tomography, such as imperfect projective measurements, affect the reconstructed states in the same way as a surviving record, and since the individual contributions of these imperfections were not quantified, the measured fidelities cannot be used to extract $V$.


\section{Discussion}\label{sec:discussion}

We have shown that a gate set of classical permutations, which is not universal even for classical computation, suffices for universal quantum computation when the instruction specifying which permutation the \F\ applies is encoded in a quantum message prepared by \W.
The permutation determines the generator, and hence the rotation axis, of each logical gate.
The rotation angle is set by \W's operation on the message outside the laboratory.
Since the \F\ reads the message only in the instruction basis, she does not have any access to the angle, and can thus gain no knowledge of it.
The erasure $E$ is required because the \F's memory stores which-path information about the executed instruction, which would otherwise suppress the interference between the instruction branches.
For a perfect erasure the message and the \F's memory return to their initial states after each gate, and only the target system is transformed.
We experimentally implemented the single-qubit sector of the protocol, for the gate family generated by $G_1$, with heralded single photons.
In particular, we used two bits to encode one logical qubit and showed that we can perform coherent rotations of this qubit around the $X$ axis.

Universal quantum computation in this framework is possible only as long as the laboratory does not leak information about which permutation the \F\ applied, implying that the computational power arises from the quantum coherence of the \F's operations, i.e. from the coherence that underlies the different descriptions that \W\ and the \F\ hold of one and the same process.
Our results thus establish an operational link between the foundations of quantum measurement and quantum computation.

Our implementation is microscopic, with the \F's memory carried by a path degree of freedom of a single photon; extending it to more complex observers would require coherent control over correspondingly larger systems.
Quantum computers may eventually allow these ideas to be tested.
For example, it has been proposed to realise the \F\ as a human-level artificial intelligence running on a quantum computer, in the context of Local Friendliness tests~\cite{wiseman2023}.
Extended Wigner's Friend circuits have already been run on quantum hardware, with friends of increasing observer-likeness quantified by a branching measure~\cite{zeng2025} and with friends implemented as reversible circuits executing simple agent-like decision rules~\cite{laux2026}, both reporting Local Friendliness violations.
Our results suggest that such experiments may also have computational consequences, since the coherent control over a friend's record that they require is the ingredient our protocol turns into universal quantum computation.
Whether this connection can be made quantitative for more complex friends remains open, as does whether the measurement of a genuine observer can be coherently reversed, which extended Wigner's Friend arguments also require.

\section{Methods}
\subsection{State preparation}\label{met:prep}

The logical input $\ket{0}_L=(\ket{10}-\ket{11})/\sqrt{2}$ is prepared from the classical state $\ket{10}$ using only the framework's allowed operations with the active instruction pair $\{\mathds{1},G_2\}$, given that the \F's action alone on the target qubits cannot generate superposition of qubits in the classical basis.

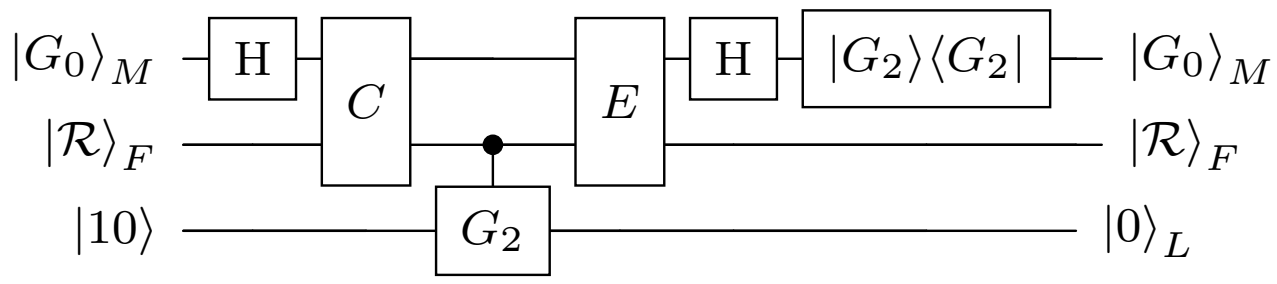
\begin{figure}[h!]
    \begin{center}
        \resizebox{\columnwidth}{!}{%
      \begin{quantikz}[column sep=0.15cm, row sep={0.5cm,between origins}, font=\footnotesize, thin lines]
        \lstick{$\ket{G_0}_M$} & \gate{\mathrm{H}} & \gate[2]{C} & \qw & \gate[2]{E} & \gate{\mathrm{H}} & \gate{\ketbra{G_2}{G_2}} & \qw& \rstick{$\ket{G_0}_M$} \\
        \lstick{$\ket{\mathcal{R}}_F$} & \qw & \qw & \ctrl{1} & \qw & \qw & \qw & \qw& \rstick{$\ket{\mathcal{R}}_F$} \\
        \lstick{$\ket{10}$} & \qw & \qw & \gate{G_2} & \qw & \qw & \qw & \rstick{$\ket{0}_L$}
      \end{quantikz}}
    \end{center}
    \caption{\textbf{State Preparation circuit.} The quantum circuit producing the logical zero $\ket{0}_L$ in the dual-rail encoding. The \F\ state, initialised in the ready state,$\ket{\mathcal{R}}_F$, is disentangled at each stage.}
    \label{fig:st_prep}
\end{figure}

\begin{enumerate}
    \item The computer is initialised in the classical state $\ket{10}$; the source produces $\ket{H}$, i.e.\ the message state $\ket{\mathds{1}}_M$.
    \item A half-wave plate (HWP) at $22.5^\circ$ implements a Hadamard on the message, preparing $(\ket{\mathds{1}}_M+\ket{G_2}_M)/\sqrt{2}\equiv\ket{D}$.
    \item The photon enters the interferometer. The entry PBS transmits the $\ket{H}$ ($=\ket{\mathds{1}}_M$) component and reflects the $\ket{V}$ ($=\ket{G_2}_M$) component, taking the \F's memory from $\ket{\mathcal{R}}_F$ to $\ket{\mathds{1}}_F$ and $\ket{G_2}_F$ respectively. In the reflected branch a calcite beam displacer shifts the occupied spatial mode vertically, $\ket{10}\mapsto\ket{11}$, executing $G_2$ (see \cref{fig:gates}{\color{bordeaux}b)}) for the experimental implementation); the transmitted path contains no optics and executes the identity. Entry is the coupling unitary $C$ of Eq.~(\ref{eq:coupling}); together with the conditional optics it realises the read--execute round of Eq.~(\ref{eq:action}).
    \item The photon exits through the second PBS, which merges the two branches into a single outside path: the \F's memory returns from $\ket{g}_F$ to $\ket{\mathcal{R}}_F$ in both branches without being read. This is the same operator applied in its erasing direction, the $E$ of Eq.~(\ref{eq:E_def}), realised optically.
    \item A further Hadamard (HWP) followed by a polariser at $90^\circ$ implements the projective measurement $\Pi_V=\ketbra{V}{V}$ on the message, leaving the state in$\ket{G}_M$ .
    \item An optional bit-flip $\hat{X}$ on the message returns it to $\ket{\mathds{1}}_M$.
\end{enumerate}

Conditioned on the projection in step 5, the computer is left in $\ket{0}_L$. 
The projection succeeds with probability $p=1/2$, and this cost originates in step 5 alone, not in the conditional gate. 
In principle the success probability can be increased by routing failed events back through the same preparation stage~\cite{horvat2022}; this recycling was not implemented here due to the architecture of the experiment.

\subsection{Tomography}\label{met:tomo}

After the protocol, the logical qubit encoded in the spatial modes is mapped onto polarisation for tomographic reconstruction, with $\ket{0}_L\equiv\ket{H}$ and $\ket{1}_L\equiv\ket{V}$. 
At the gate output the message is in $\ket{D}$, so the whole grid carries diagonal polarisation. 
A first HWP rotates the polarisation of the entire grid to $\ket{H}$, and a second HWP acting only on the second column rotates that column to $\ket{V}$:
\begin{equation}
\begin{aligned}
    &\ket{D}\otimes(\alpha\ket{0}_L+\beta\ket{1}_L)\\
    \rightarrow\;&\ket{H}\otimes(\alpha\ket{0}_L+\beta\ket{1}_L)\\
    \rightarrow\;&\alpha\ket{0}_L\otimes\ket{H}+\beta\ket{1}_L\otimes\ket{V}.
\end{aligned}
\label{eq:map}
\end{equation}
Writing the state in terms of the top ($t$) and bottom ($b$) rows of the grid,
\begin{equation}
    \tfrac{1}{\sqrt{2}}\Big((\alpha\ket{H}_t+\beta\ket{V}_t)-(\alpha\ket{H}_b+\beta\ket{V}_b)\Big),
\label{eq:row}
\end{equation}
the two rows carry identical tomographic information up to a global phase.
The columns are recombined using a calcite beam displacer.
The tomography reported in the main text is performed on the top row alone, accepting the factor-of-two loss in count rate from the $1/\sqrt{2}$ prefactor; the bottom-row cross-check below confirms the choice is ideally unbiased. 
Each tomographic basis setting is recorded for $10$~s, yielding $\approx1550$ coincidence counts. 
Density matrices $\tilde{\rho}$ are reconstructed by maximum-likelihood estimation~\cite{hradil1997quantum}, with uncertainties obtained by standard Monte Carlo resampling of the coincidence counts assuming Poissonian statistics. 
The uncertainty quoted on angle-averaged quantities is the propagated per-angle Monte Carlo uncertainty of the mean. 
The fidelity to the pure target $\sigma=\ketbra{\theta/2}{\theta/2}$, with $\ket{\theta/2}=\cos(\theta/2)\ket{0}_L-i\sin(\theta/2)\ket{1}_L$, and the Bloch vector are
\begin{equation}
\mathcal{F}=\mathrm{Tr}(\sigma\tilde{\rho}),\qquad \vec{r}=\mathrm{Tr}(\tilde{\rho}\,\vec{\sigma}),\qquad \vec{\sigma}=(\hat{X},\hat{Y},\hat{Z}).
\label{eq:fidelity}
\end{equation}

This tomography lies outside the framework of the main text. 
It is performed by the external agent directly on the \F's computer, to verify that the final state is a coherent superposition of logical basis states, which the \F, restricted to permutations, could not have produced. 
From the \F's perspective the computer holds a pair of classical bits at all times.

\section*{Acknowledgments}
We thank Clemens Lindner for helpful discussions and work on an early version of the protocol.

The financial support by the Austrian Federal Ministry of Labour and Economy, the National Foundation for Research, Technology and Development and the Christian Doppler Research Association is gratefully acknowledged.
This research was funded in whole or in part by the Austrian Science Fund (FWF) [10.55776/COE1] (Quantum Science Austria), [10.55776/F71] (BeyondC) and [10.55776/DOC162]. For open access purposes, the author has applied a CC BY public copyright license to any author accepted manuscript version arising from this submission.
Funded (Co-funded) by the European Union (HORIZON  Research and Innovation Programme, EPIQUE, No 101135288). Views and opinions expressed are however those of the author(s) only and do not necessarily reflect those of the European Union or the European Commission-EU. Neither the European Union nor the granting authority can be held responsible for them.

\section*{Data and code availability}
All the data that are necessary to replicate, verify, falsify and/or reuse this research are available online at Ref.~\cite{dataset}.
\section*{Competing interests}
The authors declare no competing interests.

\begin{widetext}
    \newpage
\begin{center}
    \textbf{Supplementary Information for: Universal Quantum Computation\\ in a Wigner's Friend Framework}
\end{center}
\appendix\label{met:met}


\section{Additional Data}\label{met:adddata}
\subsection{Tomography comparison}
\begin{figure}[h!]
    \centering
    \includegraphics[width=0.95\linewidth]{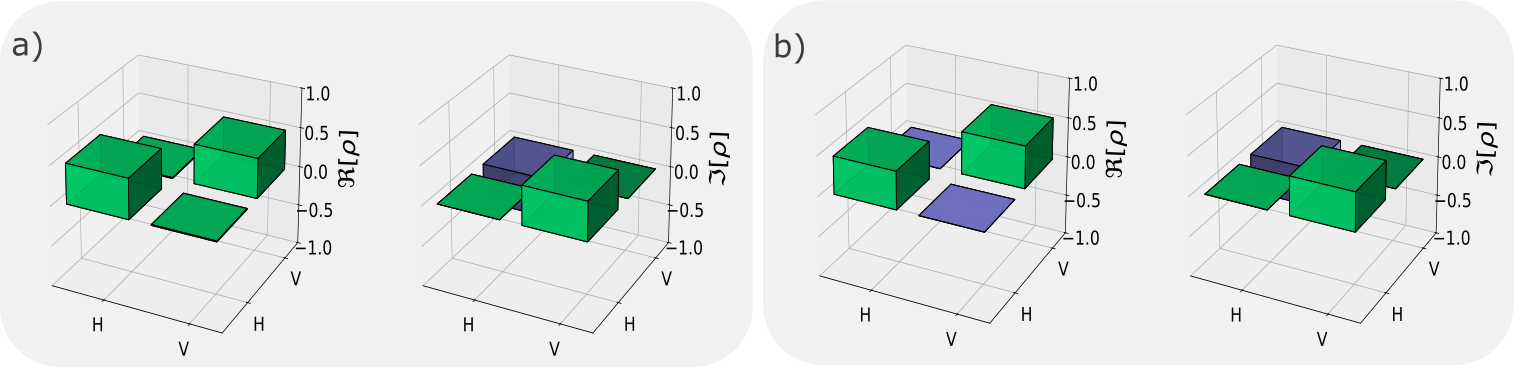}
    \caption{\textbf{Top and Bottom Tomography Comparison.} Comparison between the tomography of the top and bottom rows for $\ket{\theta/2=3\pi/4}$. In \texttt{a)} we have the tomography for the top row with $\{\mathcal{P},\mathcal{F}\}=\{0.9711(41), \, 0.9852(21)\}$. In \texttt{b)} we have the tomography of the same state using instead the bottom row. The purity and fidelity of the state is $\{\mathcal{P},\mathcal{F}\}=\{0.9714(43), \, 0.9853(22)\}$.}
    \label{fig:top_bottom}
\end{figure}
As discussed above, after the logical-to-polarisation mapping the top and bottom rows of the $2\times2$ grid carry ideally the same tomographic information up to a global phase, and the tomography reported in \cref{sec:results} uses only the top row. 
To check that this choice does not bias the result, we performed an independent tomography of the bottom row at $\theta/2 = 3\pi/4$ (corresponding to $\ket{L}=1/\sqrt{2}(\ket{H}+i\ket{V})$, up to a global phase), where the cylindrical-lens imaging system is expected to have a large effect because the output is in a maximal superposition of $\ket{0}_L$ and $\ket{1}_L$.
\cref{fig:top_bottom} shows the reconstructed density matrices for both rows.
The fidelity of the top row is $\mathcal{F}_{t} = 0.9852(21)$, and the bottom is $\mathcal{F}_{b} = 0.9853(22)$.
The two values agree within their uncertainties, supporting the use of the top row as representative of the output at the tested setting. 
The top row was chosen solely for experimental convenience.

\subsection{Single state Fidelity}
The purity $\mathcal{P} = \mathrm{Tr}(\tilde{\rho}^{2})$ and the fidelity $\mathcal{F}$ with  to the ideal target state $\ket{\theta/2}$, are computed from the same MLE-reconstructed density matrices used for the Bloch trajectory of (\cref{fig:data}). 
\cref{fig:metrics} shows the fidelity of the sixteen measured half-angles individually (plus the input state and the tomography implemented in the bottom row, discussed in \cref{met:tomo}), rather than the angle-averaged values quoted in \cref{sec:results}. The error bars come from Monte Carlo resampling of the Poissonian counts.
\begin{figure}[h!]
    \centering
    \includegraphics[width=0.75\linewidth]{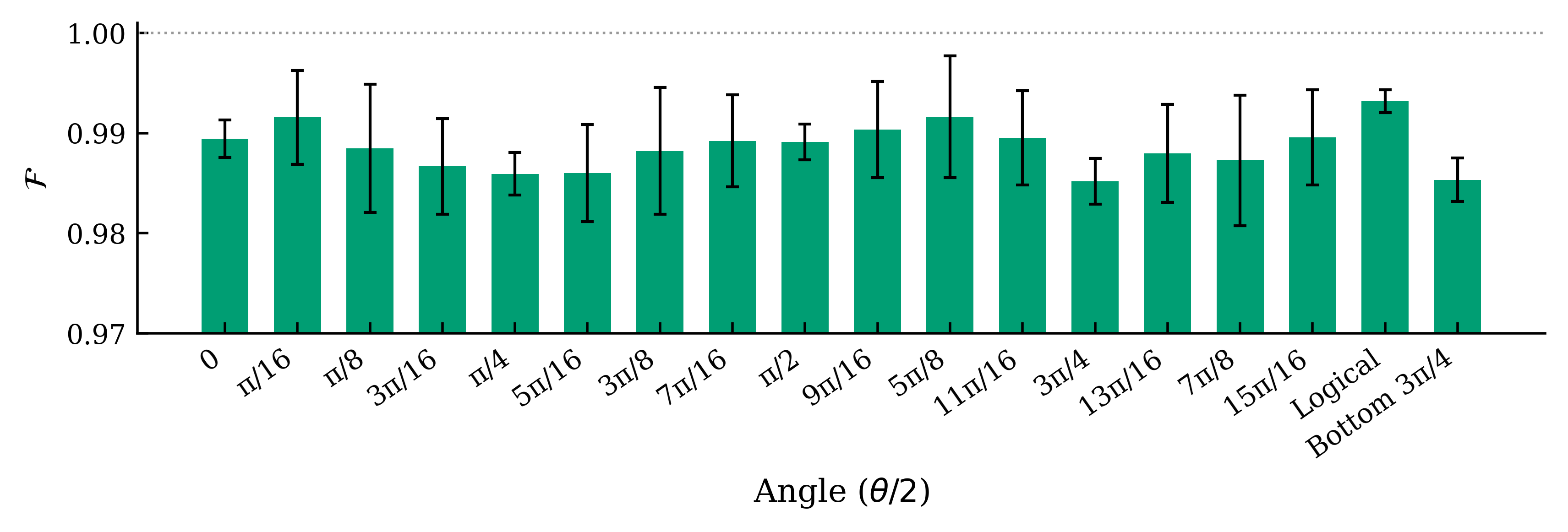}
    \caption{\textbf{Additional experimental data.} Fidelity $\mathcal{F}$ with respect to the ideal target state $|\theta/2\rangle$ (bottom), as defined in \cref{eq:fidelity}, for each of the sixteen measured half-angles, plus the input state $\ket{0}_L$, labeled as \textit{Logical}, and the state measured from the bottom row (last set of data, Bottom $3\pi/4$). Dashed lines mark the ideal value $\mathcal{F} = 1$. The average fidelity and purity values over all transformed states are $\langle\mathcal{F}\rangle = 0.9885 \pm 0.0012$ and $\langle\mathcal{P}\rangle = 0.9775 \pm 0.0023$.}
    \label{fig:metrics}
\end{figure}

\subsection{Stability}\label{sapp:stability}
\begin{figure}[h!]
    \centering
    \includegraphics[width=0.75\linewidth]{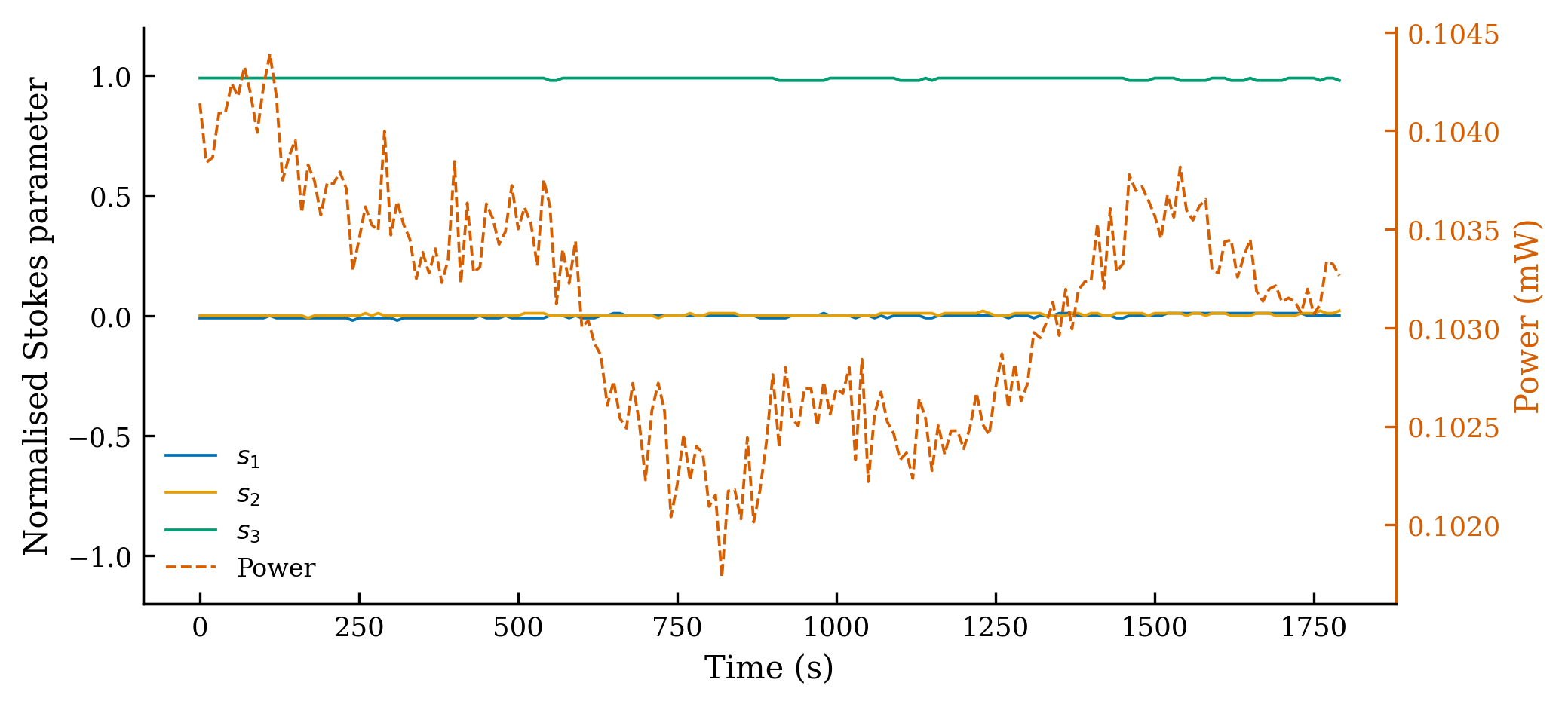}
    \caption{\textbf{Stability data.} Taken with classical light for the waveplate configuration $\theta/2 = \pi/4$, corresponding to $\ket{R}$. For the Polarimeter \texttt{PAX1000}, the convention is that $s_3=+1$ for $\ket{R}$. The plot shows the normalised Stokes parameters $s_1$, $s_2$, $s_3$, left axis and continuous lines, and the optical power , right axis and dashed line, recorded over 30 minutes. The mean recorded power was $0.1031\,\mathrm{mW}$, showing a slow drift of $2.6\%$ peak-to-peak from the laser source. The polarisation-encoded message is insensitive to this drift: the Stokes parameters remain flat to within the measurement noise, confirming that the dPBS loops and the cylindrical-lens imaging system are passively stable over a timescale longer than a full tomographic acquisition.}
    \label{fig:stability}
\end{figure}
To check that the optical setup is passively stable over the timescale of a full tomographic acquisition, we recorded the Stokes parameters and power of the output beam for $30$ minutes using classical (laser) light, with the waveplates set to the configuration that prepares the state $\ket{R}=1/\sqrt{2}(\ket{H}-i\ket{V})$ ($\theta/2 = \pi/4$). \cref{fig:stability} shows that all three Stokes parameters remain flat to within the measurement noise, with no systematic drift. Furthermore, the power also shows very small drifts over the recorded time. The dPBS paths and the cylindrical-lens $4f$ imaging system therefore remain stable for longer than a full tomographic run, and no active stabilisation is required.

\newpage
\section{Quantifying the erased record}\label{met:resource}

In the ideal protocol nothing is consumed.
\Cref{eq:final} returns the message to $\ket{+_i^{(l)}}_M$ and the memory to $\ket{\mathcal{R}}_F$ exactly.
Each round, consisting of coupling, execution and erasure, writes a record of the executed instruction into the \F's memory and then removes it.
Here we ask what happens when the removal is incomplete.

We model this with an environment onto which information about the executed instruction can leak, so that a record of the branch survives the erasure.
Before each erasure the environment acquires a partial copy of the record, $\ket{g}_F\ket{0}_E\to\ket{g}_F\left(\cos\eta\ket{0}+\sin\eta\ket{g}\right)_E$ for $g\in\{\mathds{1},G\}$, with $\ket{\mathcal{R}}_F$ unaffected.
For $\eta=0$ nothing leaks, and for $\eta=\pi/2$ a perfect copy of the record survives.
In this model each environment has the Hilbert space
\begin{equation}
    \mathcal{H}^{(i)}_E=\mathrm{span}\left(\{\ket{0}_E\}\cup\left\{\ket{g}_E : g\in\{\mathds{1},G\}\right\}\right),
    \label{eq:he}
\end{equation}
where the superscript labels the leak, $i=1$ before the first erasure and $i=2$ before the second.
We will see that the second leak does not affect the target.
We write $\ket{\cdot}_E$ for states in $\mathcal{H}^{(1)}_E$ and $\ket{\cdot}_{E_2}$ for states in $\mathcal{H}^{(2)}_E$.

The model assumes that the leak acts only on the \F's memory, between execution and erasure, and leaves the message and the target untouched, and that both branches leak with the same strength $\eta$ into orthogonal environment states.
It only covers photons that are detected.
Photons that leave through the unused output port of the exit dPBS are lost before detection, so they lower the count rate but not the measured fidelity.

We abbreviate the environment states as $\ket{e_g}_E=\cos\eta\ket{0}_E+\sin\eta\ket{g}_E$ and denote by $V$ the real part of their overlap,
\begin{equation}
    V=\mathrm{Re}\,\braket{e_{\mathds{1}}|e_G}_E=\cos^2\eta .
    \label{eq:res}
\end{equation}
In this model $V=1$ means that the two records are identical and $V=0$ that they are perfectly distinguishable, so $1-V$ quantifies how distinguishable the records are.

Including the environment, the protocol runs as follows.
\begin{enumerate}
    \item The initial state is
    $$\frac{1}{\sqrt{2}}\left(\ket{\mathds{1}}_M+\ket{G}_M\right)\ket{\mathcal{R}}_F\ket{\psi}_L\otimes\ket{0}_E\ket{0}_{E_2}\,.$$
    \item The coupling and execution step gives
    $$\frac{1}{\sqrt{2}}\left(\ket{\mathds{1}}_M\ket{\mathds{1}}_F\ket{\psi}_L+\ket{G}_M\ket{G}_F\,G\ket{\psi}_L\right)\otimes\ket{0}_E\ket{0}_{E_2}\,.$$
    \item The first leak copies part of the record into the environment,
    $$\frac{1}{\sqrt{2}}\left(\ket{\mathds{1}}_M\ket{\mathds{1}}_F\ket{e_{\mathds{1}}}_E\ket{\psi}_L+\ket{G}_M\ket{G}_F\ket{e_G}_E\,G\ket{\psi}_L\right)\otimes\ket{0}_{E_2}\,.$$
    \item \W\ applies the erasure,
    $$\frac{1}{\sqrt{2}}\left(\ket{\mathds{1}}_M\ket{e_{\mathds{1}}}_E\ket{\psi}_L+\ket{G}_M\ket{e_G}_E\,G\ket{\psi}_L\right)\otimes\ket{\mathcal{R}}_F\ket{0}_{E_2}\,.$$
    \item \W\ rotates the message. With $c_\theta=\cos\frac{\theta}{2}$ and $s_\theta=\sin\frac{\theta}{2}$, the state becomes
    $$\frac{1}{\sqrt{2}}\Big[\ket{\mathds{1}}_M\big(c_\theta\ket{e_{\mathds{1}}}_E\ket{\psi}_L-is_\theta\ket{e_G}_E\,G\ket{\psi}_L\big)+\ket{G}_M\big(c_\theta\ket{e_G}_E\,G\ket{\psi}_L-is_\theta\ket{e_{\mathds{1}}}_E\ket{\psi}_L\big)\Big]\otimes\ket{\mathcal{R}}_F\ket{0}_{E_2}\,.$$
    \item The second coupling and execution follow. Writing $\bar g$ for the label of the other branch and defining the combined environment--target states $\ket{w_g}_{EL}=c_\theta\ket{e_g}_E\ket{\psi}_L-is_\theta\ket{e_{\bar g}}_E\,G\ket{\psi}_L$, the state is
    $$\frac{1}{\sqrt{2}}\left(\ket{\mathds{1}}_M\ket{\mathds{1}}_F\ket{w_{\mathds{1}}}_{EL}+\ket{G}_M\ket{G}_F\ket{w_G}_{EL}\right)\otimes\ket{0}_{E_2}\,.$$
    \item The second leak gives
    $$\frac{1}{\sqrt{2}}\left(\ket{\mathds{1}}_M\ket{\mathds{1}}_F\ket{e_{\mathds{1}}}_{E_2}\ket{w_{\mathds{1}}}_{EL}+\ket{G}_M\ket{G}_F\ket{e_G}_{E_2}\ket{w_G}_{EL}\right).$$
    \item \W\ applies the second erasure, and the final state is
    $$\frac{1}{\sqrt{2}}\left(\ket{\mathds{1}}_M\ket{e_{\mathds{1}}}_{E_2}\ket{w_{\mathds{1}}}_{EL}+\ket{G}_M\ket{e_G}_{E_2}\ket{w_G}_{EL}\right)\otimes\ket{\mathcal{R}}_F\,.$$
\end{enumerate}

To obtain the state of the target we trace out the message, both environments and the \F.
Since $\ket{\mathds{1}}_M$ and $\ket{G}_M$ are orthogonal, the trace over $M$ removes every cross-term between the two branches.
Tracing out $M$, and dropping the decoupled memory $F$, gives
$$\rho_{LE,E_2}=\frac{1}{2}\Big(\ket{e_{\mathds{1}}}_{E_2}\!\bra{e_{\mathds{1}}}\otimes\ket{w_{\mathds{1}}}_{EL}\!\bra{w_{\mathds{1}}}+\ket{e_G}_{E_2}\!\bra{e_G}\otimes\ket{w_G}_{EL}\!\bra{w_G}\Big).$$
The trace over the second environment only uses the normalisation $\braket{e_g|e_g}_{E_2}=1$.
With the cross-terms already gone, the overlap $\braket{e_{\mathds{1}}|e_G}_{E_2}$ never appears, and the second leak drops out of the target's dynamics,
$$\rho_{LE}=\frac{1}{2}\Big(\ket{w_{\mathds{1}}}_{EL}\!\bra{w_{\mathds{1}}}+\ket{w_G}_{EL}\!\bra{w_G}\Big).$$
Tracing over the first environment gives $\rho_L=\frac{1}{2}(T_{\mathds{1}}+T_G)$ with $T_g=\mathrm{Tr}_E\big[\ket{w_g}\!\bra{w_g}_{EL}\big]$.
Expanding $\ket{w_g}_{EL}$, with $\rho=\ket{\psi}_L\!\bra{\psi}$,
\begin{align*}
    T_{\mathds{1}}&=c_\theta^2\rho+s_\theta^2\,G\rho G+ic_\theta s_\theta\braket{e_G|e_{\mathds{1}}}_E\,\rho G-ic_\theta s_\theta\braket{e_{\mathds{1}}|e_G}_E\,G\rho\,,\\
    T_G&=c_\theta^2\rho+s_\theta^2\,G\rho G+ic_\theta s_\theta\braket{e_{\mathds{1}}|e_G}_E\,\rho G-ic_\theta s_\theta\braket{e_G|e_{\mathds{1}}}_E\,G\rho\,.
\end{align*}
In the sum each overlap appears together with its complex conjugate, so the imaginary parts cancel and only the real part survives,
\begin{equation}
    \rho_L=c_\theta^2\rho+s_\theta^2\,G\rho G+ic_\theta s_\theta V\left(\rho G-G\rho\right).
    \label{eq:imp_state}
\end{equation}
The derivation uses the two environment states only through their overlap, so \cref{eq:imp_state} holds for any pure records $\ket{e_{\mathds{1}}}_E$ and $\ket{e_G}_E$, not only for the model of \cref{eq:he}.
In particular it covers a record that carries a relative phase, $\braket{e_{\mathds{1}}|e_G}_E=|\braket{e_{\mathds{1}}|e_G}_E|\,e^{i\chi}$, for which $V=|\braket{e_{\mathds{1}}|e_G}_E|\cos\chi$, and a dephased record and a partially copied record act identically on the target whenever they share the same $V$.

To identify the channel in \cref{eq:imp_state}, we compare it with a mixture of the two rotation signs.
Writing $U_{\pm\theta}=c_\theta\mathds{1}\mp is_\theta G$ and expanding,
$$U_{\theta}\rho U_{\theta}^{\dagger}=c_\theta^2\rho+s_\theta^2\,G\rho G+ic_\theta s_\theta\left(\rho G-G\rho\right),\qquad U_{-\theta}\rho U_{-\theta}^{\dagger}=c_\theta^2\rho+s_\theta^2\,G\rho G-ic_\theta s_\theta\left(\rho G-G\rho\right),$$
so that for any weight $p$
$$(1-p)\,U_{\theta}\rho U_{\theta}^{\dagger}+p\,U_{-\theta}\rho U_{-\theta}^{\dagger}=c_\theta^2\rho+s_\theta^2\,G\rho G+ic_\theta s_\theta(1-2p)\left(\rho G-G\rho\right).$$
This reproduces \cref{eq:imp_state} term by term if and only if $1-2p=V$, and the target therefore evolves as
\begin{equation}
    \Lambda(\rho)=(1-p)\,U_{\theta}\rho U_{\theta}^{\dagger}+p\,U_{-\theta}\rho U_{-\theta}^{\dagger},\qquad p=\frac{1-V}{2}.
    \label{eq:pm_channel}
\end{equation}
The overlap multiplies only the commutator term $\rho G-G\rho$, which is the part of $\rho_L$ that changes sign with $\theta$.
The coherence of the record therefore carries the phase relation that distinguishes $U_\theta$ from $U_{-\theta}$, and losing it randomises the sense of the rotation while leaving its size intact.
For $V=0$, reached for instance by a perfect copy ($\eta=\pi/2$), the channel is an equal mixture of $U_\theta$ and $U_{-\theta}$.

For an input $\ket{\psi}$ with pure target $\ket{\phi}=U_\theta\ket{\psi}$, the fidelity of the output is $\mathcal{F}=\bra{\phi}\Lambda(\ket{\psi}\!\bra{\psi})\ket{\phi}$.
With \cref{eq:pm_channel} this gives $\mathcal{F}=(1-p)+p\,|\bra{\psi}U_\theta^\dagger U_{-\theta}\ket{\psi}|^2$, where the second term is the squared overlap between the ideal output and the output rotated in the opposite direction.
Using $G^2=\mathds{1}$, $U_\theta^\dagger U_{-\theta}=(c_\theta\mathds{1}+is_\theta G)^2=\cos\theta\,\mathds{1}+i\sin\theta\,G=e^{i\theta G}$.
With $\tilde g=\bra{\psi}G\ket{\psi}$, which is real since $G$ is Hermitian, $|\bra{\psi}e^{i\theta G}\ket{\psi}|^2=\cos^2\theta+\tilde g^{2}\sin^2\theta$, so that
\begin{equation}
    1-\mathcal{F}=p\,(1-\tilde g^{2})\sin^2\theta .
    \label{eq:general_law}
\end{equation}
The input $\ket{0}_L$ used in the experiment has $\tilde g=\bra{0}\hat X\ket{0}=0$, giving $1-\mathcal{F}(\theta)=\frac{1-V}{2}\sin^2\theta$.
Eigenstates of $G$, with $\tilde g=\pm1$, are taken to the same state by both rotations up to a global phase, so this mechanism leaves their fidelity unchanged.
An input orthogonal to the rotation axis, such as $\ket{0}_L$, is instead maximally sensitive, since the two rotation signs send it to the most distant pair of states.

To average \cref{eq:general_law} over input states it is enough to average $\tilde g^{2}$.
Since $G$ is Hermitian and unitary with $G^2=\mathds{1}$, on a single logical qubit it has eigenvalues $\pm1$ and vanishing trace, so it is a Pauli operator along some axis, $G=\hat u\cdot\vec\sigma$, with $\hat u$ a real unit vector and $\vec\sigma=(\hat X,\hat Y,\hat Z)$.
Writing a pure input as $\ket{\psi}\!\bra{\psi}=\frac{1}{2}(\mathds{1}+\hat n\cdot\vec\sigma)$ with Bloch vector $\hat n$, and using $\mathrm{Tr}(\sigma_i\sigma_j)=2\delta_{ij}$ together with $\mathrm{Tr}\,\sigma_i=0$,
\begin{equation}
    \tilde g=\bra{\psi}G\ket{\psi}=\tfrac{1}{2}\,\mathrm{Tr}\big[(\mathds{1}+\hat n\cdot\vec\sigma)(\hat u\cdot\vec\sigma)\big]=\hat n\cdot\hat u=\cos\vartheta ,
    \label{eq:g_geom}
\end{equation}
where $\vartheta$ is the angle between the input Bloch vector and the axis of $G$.
The overlap $\tilde g$ is thus the projection of the input onto the axis of the operation the \F\ performs.
Haar-random pure states of a qubit have Bloch vectors distributed uniformly on the sphere, with normalised measure $d\mu=\sin\vartheta\,d\vartheta\,d\phi/4\pi$, so with $u=\cos\vartheta$
\begin{equation}
    \langle\tilde g^{2}\rangle=\frac{1}{4\pi}\int_{0}^{2\pi}\!d\phi\int_{0}^{\pi}\!\cos^{2}\vartheta\,\sin\vartheta\,d\vartheta=\frac{1}{2}\int_{-1}^{1}u^{2}\,du=\frac{1}{3},
    \label{eq:g2_haar}
\end{equation}
independently of the axis $\hat u$.
Substituting into \cref{eq:general_law} gives $1-\bar{\mathcal{F}}(\theta)=\frac{1-V}{3}\sin^2\theta$, the input-independent statement about the channel.
The experiment measures the state-specific law with the coefficient $\frac{1}{2}$, since it prepares $\ket{0}_L$ rather than an ensemble of inputs.
The factor $\sin^2\theta$ vanishes at $\theta=0$ and $\theta=\pi$, where the gate is a permutation up to a global phase and needs no interference between the instruction branches, and is largest at $\theta=\pi/2$ (\cref{fig:resource_law}).

The second leak attaches its record to the branch label, so the overlap $\braket{e_{\mathds{1}}|e_G}_{E_2}$ can only multiply cross-terms between the two branches, and those have already been removed by the orthogonality of the untouched message states.
The first leak behaves differently because \W's rotation in step 5 mixes the two branches, placing both $\ket{e_{\mathds{1}}}_E$ and $\ket{e_G}_E$ inside each $\ket{w_g}_{EL}$, where they are correlated with the target rather than with the branch label.
The role of the second erasure is to return the message and the memory to their initial states, so an imperfect second erasure degrades subsequent rounds rather than the present one.
In the ideal case, with no leak, the second read--execute round maps the state to $\frac{1}{\sqrt{2}}\left(\ket{\mathds{1},\mathds{1}}+\ket{G,G}\right)\otimes U_\theta\ket{\psi}$, using $G^2=\mathds{1}$, and the computer has already factored out with the gate complete before the erasure is applied.\\

\noindent\emph{Experimental imperfections.}
In the experiment the \F's record is the path of the photon through the interferometer, so any degree of freedom that still distinguishes the two arms after the exit dPBS is a surviving record in the sense of \cref{eq:imp_state}.
Aberrations of the cylindrical-lens telescope, which sits only in the reflected arm, can leave the two arms in slightly different transverse modes, so that the path remains partly distinguishable after the exit dPBS.
Polarisation mixing in the dPBS makes the coupling and erasure themselves imperfect, so that the two arms are not recombined exactly.
The tomography contributes separately through the finite contrast of each measurement basis, which, as shown below, affects the reconstructed states in the same way as a surviving record.\\

\noindent\emph{Comparison with the data.}
The measured infidelities of \cref{sec:results} also contain contributions from technical imperfections, and the angular dependence alone cannot separate them from a surviving record.
To see this, write the ideal output as $\ket{\theta/2}=\cos\frac{\theta}{2}\ket{0}_L-i\sin\frac{\theta}{2}\ket{1}_L$, whose Bloch vector is
\begin{equation}
    \vec t(\theta)=\left(0,\,-\sin\theta,\,\cos\theta\right).
    \label{eq:tgt_bloch}
\end{equation}
For any reconstructed state $\tilde\rho=\frac{1}{2}(\mathds{1}+\vec r\cdot\vec\sigma)$ the fidelity to this pure target is $\mathcal{F}=\frac{1}{2}(1+\vec t\cdot\vec r)$, so
\begin{equation}
    1-\mathcal{F}=\tfrac{1}{2}\left(1+r_Y\sin\theta-r_Z\cos\theta\right),
    \label{eq:fid_bloch}
\end{equation}
and only the $Y$ and $Z$ components of the output enter.
Applied to $\ket{0}_L$, the channel of \cref{eq:pm_channel} mixes the Bloch vectors $(0,\mp\sin\theta,\cos\theta)$ with weights $1-p$ and $p$, giving $\vec r=(0,-V\sin\theta,\cos\theta)$: the leak leaves the $Z$ component unchanged and multiplies the $Y$ component by $V$.
A tomography with finite contrast $c_k\leq1$ in the basis $k$ has measurement operators $\frac{1}{2}(\mathds{1}\pm c_k\sigma_k)$ and returns $c_k r_k$ in place of each Bloch component, as long as the reconstruction lies inside the Bloch ball.
With both effects present the reconstructed state has $\vec r=(0,-c_YV\sin\theta,c_Z\cos\theta)$, and \cref{eq:fid_bloch} gives
\begin{equation}
    1-\mathcal{F}=\frac{1-c_Z}{2}+\frac{c_Z-c_YV}{2}\,\sin^2\theta .
    \label{eq:readout_infid}
\end{equation}
For $c_Y=c_Z=1$ this reduces to \cref{eq:general_law}, but a readout that is slightly worse in the $Y$ basis than in the $Z$ basis produces the same $\sin^2\theta$ dependence with no surviving record at all.
More generally, $V$ and $c_Y$ enter the reconstructed state only through their product $c_YV$, so the pairs $(c_Y,V)$ and $(c_YV,1)$ give identical density matrices at every angle.
A fit of the per-angle infidelities fixes $c_Z$ and $c_YV$ but cannot separate the two factors, and we therefore treat \cref{eq:general_law} as a prediction of the framework rather than extracting $V$ from the data.

\begin{figure}[htbp]
    \centering
    \includegraphics[width=0.55\linewidth]{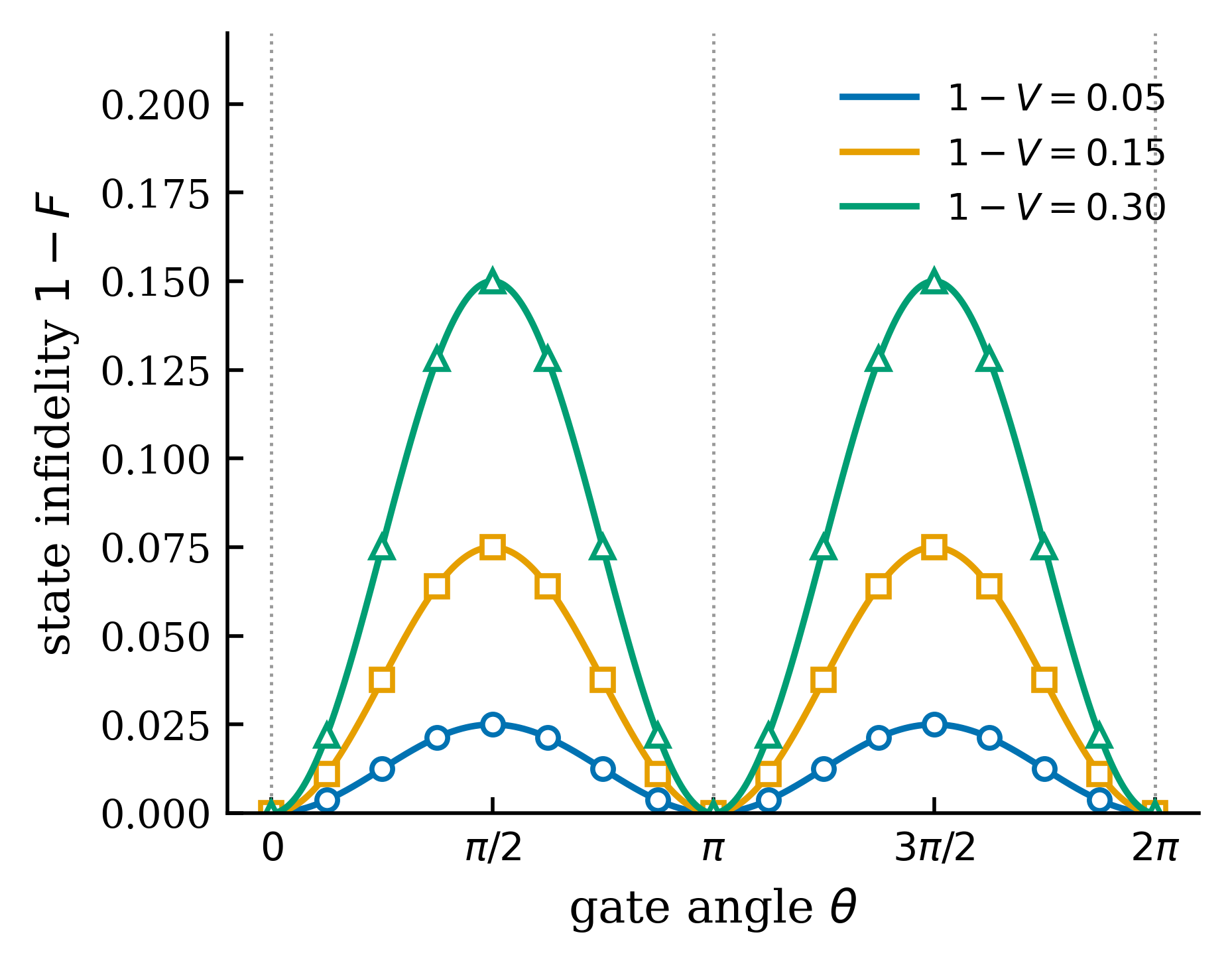}
    \caption{\textbf{The erasure--fidelity law.} State infidelity of the gate as a function of $\theta$ for three values of $1-V$. Lines are the closed form $\frac{1-V}{2}\sin^2\theta$ of \cref{eq:general_law} for the input $\ket{0}_L$; markers are the exact simulation of the six-step protocol with the leak model. Dotted lines mark $\theta=0,\pi,2\pi$, where the infidelity vanishes because the gate is a permutation up to a global phase.}
    \label{fig:resource_law}
\end{figure}
\end{widetext}
\end{document}